%% file: Main_text_incl._figures.tex
\documentclass[%
 reprint,
 amsmath,amssymb,
 aps,
]{revtex4-1}

\usepackage{graphicx}
\usepackage{dcolumn}
\usepackage{bm}
\usepackage{verbatim}
\usepackage{color}
\usepackage{gensymb}
\usepackage[version=3]{mhchem}
\usepackage[usenames,dvipsnames]{xcolor}
\definecolor{dark-green}{HTML}{006400}
\usepackage{threeparttable}
\usepackage{adjustbox}
\usepackage{footnote}
\usepackage{float}
\usepackage{hyperref}
\usepackage{scrextend}
\newcommand{\lro}{Li$_2$RuO$_3$}

\begin{document}

\preprint{APS/123-QED}

\title{Effect of disorder on the dimer transition of the honeycomb-lattice compound Li$_{2}$RuO$_3$}

\author{Marco-Polo Jimenez-Segura}
\author{Atsutoshi Ikeda}%
\author{Shingo Yonezawa}%
\author{Yoshiteru Maeno}%
\affiliation{%
 Department of Physics, Graduate School of Science, Kyoto University, Kyoto 606-8502, Japan\\
 }%

\date{\today}

\begin{abstract}
We report the dependence of magnetic properties on the crystalline disorder in Li$_{2}$RuO$_3$ with Ru honeycomb lattice. This oxide exhibits unconventional Ru-dimer transition below $T_\text{d} \sim$ 540 K. We demonstrate that the cell parameters, related to the coherence of the dimer formation, are strongly dependent on the synthesis procedure. We show that the magnetic behavior at the dimer transition is closely related to the lattice parameters. In particular, we revealed that samples with well-ordered dimers exhibit a first-order magnetic transition with the onset exceeding 550~K, higher than that reported previously.
We discuss possible dimer configurations leading to this magneto-lattice coupling.
\end{abstract}

\pacs{Valid PACS appear here}
\maketitle
\section{\label{Introduction}Introduction}
Honeycomb lattices with magnetic ions at the vertices of the hexagon have been under extensive experimental and theoretical studies \cite{kitaev_anyons_2006, schaffer_quantum_2012, Chaloupka_zigzag_2013, trousselet_hidden_2013, pchelkina_electronic_2015, grandi_topological_2015}. One of the main motivations is to find topological superconductivity predicted to emerge with hole doping in the Kitaev-Heisenberg model with the spin \textit{S} = 1/2  \cite{you_doping_2012, hyart_competition_2012, scherer_unconventional_2014, okamoto_global_2013}.
Some examples that can be modeled with the Kitaev-Heisenberg Hamiltonian are \textit{A}$_2$IrO$_3$ (\textit{A} = Li, Na). However, experimental evidence of superconductivity has not been reported yet \cite{singh_antiferromagnetic_2010, kobayashi_structure_2003, felner_magnetic_2002, takayama_hyperhoneycomb_2015, singh_relevance_2012, singh_antiferromagnetic_2010}. More recently, through substitution of Ir$^{4+}$ ($5d^5$) by Ru$^{4+}$ ($4d^4$) in Li$_2$IrO$_3$, the compound \textit{A}$_2$Ir$_{1-x}$Ru$_{x}$O$_3$ (\textit{A} = Li, Na) is found to be even non-metallic \cite{lei_structural_2014}.

\begin{figure}[t]
\includegraphics[width=9cm]{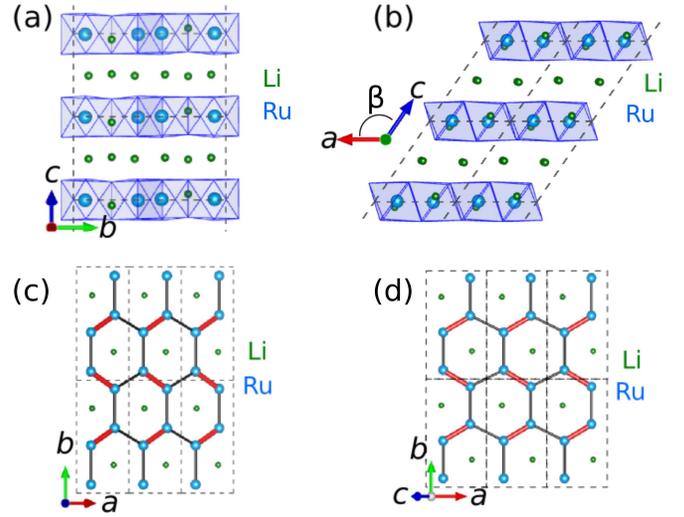}
\caption{\label{cell_init} (Color online) Room-temperature structure of Li$_2$RuO$_3$ with the space group $P2_1/m$ viewed from several directions. Blue and green spheres represent Ru and Li ions, while vertices of the octahedra represent O ions. Black and red lines represent the long and short bonds whithin the Ru honeycomb lattice. Panels (a), (b), and (c) show the structure as viewed along the \textit{a}, \textit{b} and \textit{c} directions, while (d) is along \textit{c}*, the direction perpendicular to the \textit{ab} plane. The figures are prepared with the program VESTA \cite{momma_vesta_2011}.}
\end{figure}

Other examples of honeycomb-lattice compounds are \textit{A}$_2$RuO$_3$ (\textit{A} = Li, Na) \cite{james_structure_1988, felner_magnetic_2002, dulac_synthese_1970, kimber_interlayer_2010} with nominally \textit{S}=1 originating from Ru$^{4+}$ in the low-spin state. In contrast to \textit{A}$_2$IrO$_3$ (\textit{A} = Li, Na), it has been discovered that  Li$_2$RuO$_3$ exhibits an unusual phase transition at $T_{\text{d}}~\sim~ 540~\text{K}$ \cite{miura_new-type_2007}. Below $T_{\text{d}} $, two of the six Ru-Ru bonds in the honeycomb hexagon become substantially shorter than the others, forming static Ru-Ru dimers, as schematically shown with thick red lines in Fig.$~$\ref{cell_init}. Because of this static dimerization, the high-temperature structure with a nearly ideal honeycomb lattice, belonging to the space group $C2/m$, is reduced to a less symmetric structure with distorted honeycomb lattice belonging to the space group $P2_1/m$ below $T_{\text{d}}$ \cite{miura_new-type_2007, lei_structural_2014, kimber_valence_2014}. Interestingly, this dimer transition is accompanied by a strong decrease of the magnetization. In order to explain the origin of this transition, scenarios such as the transition from a highly correlated metal to a molecular-orbital insulator accompanied by bond-dimer formation \cite{miura_new-type_2007, miura_structural_2009} and the formation of spinless dimers by magneto-elastic mechanism \cite{jackeli_classical_2008} have been proposed. More recently, a pair distribution function (PDF) analysis based on high-energy X-ray diffraction (XRD) revealed that the dimers exist even above $T_{\text{d}}$ but the positions of the dimers change dynamically \cite{kimber_valence_2014}. Thus, the transition at $T_\text{d}$ can be regarded as the ``melting" transition between the dimer-liquid and dimer-solid phases.
 
In this work, we study the correlation between the cell parameters of Li$_{2}$RuO$_3$ and the magnetic behavior of the dimer transition, focusing on the importance of the crystallinity for the observation of intrinsic properties of Li$_2$RuO$_3$. 

\section{\label{EXPERIMENT}EXPERIMENT}
Polycrystalline Li$_2$RuO$_3$ samples were prepared from Li$_2$CO$_3$ (Aldrich, 99.997\%) and RuO$_2$ (Rare Metallic, 99.9\%). Li$_2$CO$_3$ was dried at 300\celsius\ for 2 hours. Their masses were measured, and then they were mixed and ground for 1 hour in a conventional mortar. This starting powder was pelletized and heated for 24 hours as the first heating process. The process of grinding, pelletizing and heating was repeated several times for some samples. Details of the synthesis are summarized in Table~\ref{cellpar}. 
 
Powder XRD measurements were performed at room temperature with a commercial diffractometer (Bruker AXS, D8 Advance) using the CuK$_{\alpha}$ radiation equipped with a one-dimensional array of detectors and a nickel monochromator. The sample stage was spun at 30 rpm. Peak indexing was  carried out using the program Topas 4.2. The tube tails as well as other instrument parameters were calibrated using a standard reference plate of alumina provided by the National Institute of Standards and Technology.
Magnetization measurements were carried out using a commercial superconducting quantum interference device (SQUID) magnetometer (Quantum Design, MPMS). Measurements of magnetization at high temperatures (300 to 700 K) were performed using the oven option for MPMS. We confirmed the validity of the calibration of the oven thermometer by measuring the ferromagnetic transition ($T_\text{C}=627.2$~K \cite{kouvel_detailed_1964, arajs_critical_1970}) of Ni (Rare Metallic, 99.99\%) at several fields. The calibration error of the thermometer at $T_{\text{C}}$ of Ni was found to be less than 0.2\%. 
 
\section{\label{RESULTS AND DISCUSSION}RESULTS AND DISCUSSION}
\subsection{\label{structure} Crystalline structure}
\begin{figure}[t]
\includegraphics[width=9cm]{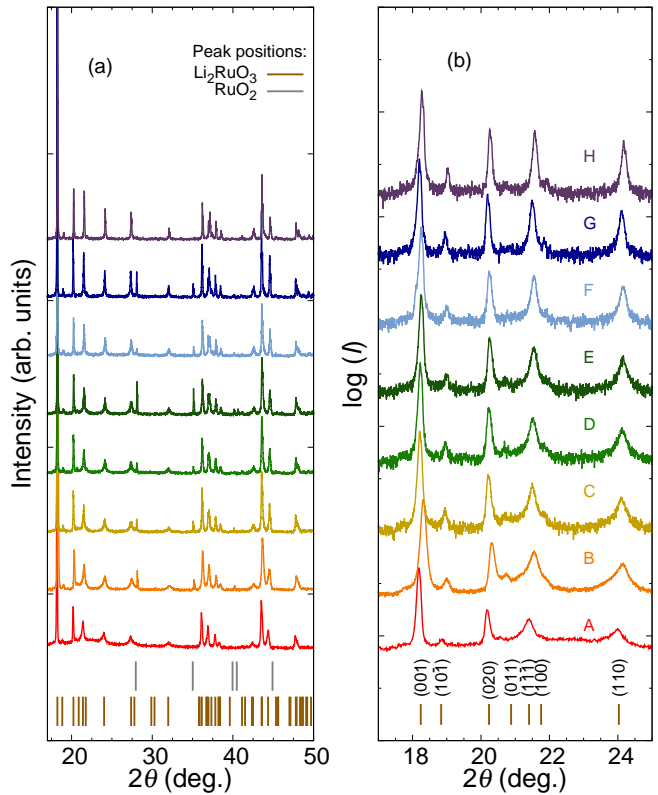}
\caption{\label{XRD_pristine} (Color online) Comparison of XRD spectra of several samples of Li$_2$RuO$_3$. The labels A--H identify samples listed in Table~\ref{cellpar}. Small horizontal shifts of the spectra mainly arisen from small variation in the position of the sample surface are observed. This shift was corrected when the cell parameters were obtained. The panel (a) shows the intensity \textit{I} vs 2$\theta$ and (b) the semi-logarithmic intensity \textit{I} vs 2$\theta$ in a narrower angle range. Expected peak positions for Li$_2$RuO$_3$ ($P2_1/m$) and RuO$_2$ are indicated with brown and gray vertical lines.}
\end{figure}
\begin{figure}[t]
\includegraphics[width=9cm]{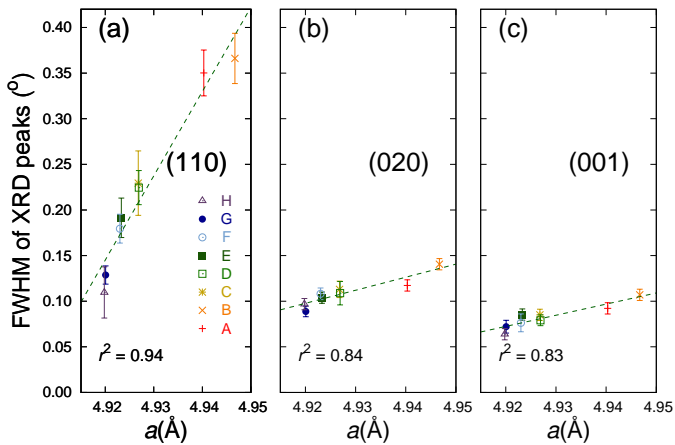}
\caption{\label{XRD_FWHM}
(Color online) Full width half maximum (FWHM) of the peaks (110), (020) and (001) of the XRD peaks as a function of the cell parameter \textit{a}, shown in the panels (a), (b), and (c) respectively.}
\end{figure}

\begin{table*}[t]
\caption{\label{cellpar}%
Summary of lattice parameters of Li$_{2}$RuO$_3$ and synthesis conditions.}
\begin{ruledtabular}
\begin{center}
\begin{tabular}{cccccccccc}  
\shortstack{sample\\ label \\  \textcolor{white}{.} \\  \textcolor{white}{.}  \\  \textcolor{white}{.}} & 
\shortstack{starting \\ composition \\  Li : Ru \\(mol : mol) \\ \textcolor{white}{.}} & 
\shortstack{Max. \\ temp.\\of the 1st \\ heating step\footnote{For 24 hours.}\footnote{We spent 2.5 hours from room temperature to $T_{\text{max}}$.\label{fn:heating}}\\ ($^{\circ}$C)}& 
\shortstack{Max. \\ temp.\\of the 2nd \\ heating step\footref{fn:heating}\footnote{For 48 hours.\label{fn:second}}\\ ($^{\circ}$C)}&
\shortstack{Max. \\ temp.\\of the 3rd \\ heating step \footref{fn:heating}\footref{fn:second}\\ ($^{\circ}$C)}&
\shortstack{ \textit{a} (\AA) \textcolor{white}{.}  \\  \textcolor{white}{.} \\  \textcolor{white}{.} } &
\shortstack{ \textit{b} (\AA) \textcolor{white}{.}  \\  \textcolor{white}{.} \\  \textcolor{white}{.} } &
\shortstack{ \textit{c} (\AA) \textcolor{white}{.}  \\  \textcolor{white}{.}  \\  \textcolor{white}{.}} &
\shortstack{ $\beta$ ($^{\circ}$) \textcolor{white}{.}  \\  \textcolor{white}{.} \\  \textcolor{white}{.} }
  \\
\colrule
A & 2.08 : 1 & 1000 & - & - & 4.9403(13) & 8.7655(23) & 5.8893(17) & 124.4395(40) 

\\ 
B & 2 : 1 & 1000 & -& - & 4.9470(05) &	 8.7622(09) & 5.8916(06) &	124.4615(29)  

 \\ 
C & 2 : 1 & 1000 & 900 & - & 4.9268(10) &	8.7755(18) & 5.8958(13) & 124.3760(31) 	

\\
D & 2 : 1 & 1000 & 900 & 900 & 4.9269(13) & 8.7773(23) & 5.8974(16) &	 124.3757(30) 

\\
E & 2 : 1 & 1000 & 900 & 900 & 4.9233(09)	 & 8.7741(15) & 5.8941(11) & 124.3726(31) 	

\\
F & 2 : 1 & 1000 & 900 & 1000 & 4.9230(08)	 & 8.7806(14)	& 5.8959(10) &	124.3610(23) 		

\\
G & 2 : 1 & 1000 & 1000 & - &  4.9200(04)	& 8.7809(08) &	5.8940(06)	& 124.3501(14)	

\\
H\footnote{Acetone was added while grinding in order to enhance the homogeneity.} & 2 : 1 & 1000 & 1000 & - & 4.9198(04)	 & 8.7822(07) & 5.8932(05) &	124.3485(13) 

\\
\end{tabular}
\end{center}

\end{ruledtabular}
\end{table*}

In this study, we compare results of 8 samples. The characters of the samples can be divided by their synthesis procedures (Table \ref{cellpar}): samples A and B were heated only once at 1000\celsius\ \cite{miura_new-type_2007}; samples C, D, and E were heated two or more times with final heating at 900\celsius\ \cite{james_structure_1988}; samples F to H were heated two or more times with final heating at 1000\celsius . All samples were furnace-cooled after the heater was switched off. In addition, in order to improve the homogeneity of the starting powder for sample H, acetone was added to the powder during the grinding. 

In Fig.~\ref{XRD_pristine}, powder XRD spectra of samples A to H are presented. All the spectra are well fitted with the reported space group $P2_1/m$ with the static Ru-Ru dimers \cite{miura_new-type_2007}. Samples A and H do not show any secondary phase by XRD, while the other samples contain a small amount (at most 4\%) of RuO$_2$.

Surprisingly, we found that the cell parameters and crystallinity of the samples are strongly affected by the synthesis process. In Fig.~\ref{XRD_pristine}(b), spectra between 17$^\circ$ and 25$^\circ$ are presented with a logarithmic vertical scale. It is clear that peaks become sharper for samples with labels in the order of the alphabet (A to H). In particular, peaks related to the \textit{a} parameter, i.e. (\textit{hkl}) peaks with $h\neq 0$, exhibit stronger sample dependence. For instance, the ($10\bar{1}$) and (110) peaks have evidently different sharpness depending on the samples and these peaks are significantly broader for samples A and B, which were heated only once. In contrast, the peaks belonging to directions that only include the \textit{c} direction (i.e. (00\textit{l}) peaks) exhibit a weaker change. 

For a more quantitative analysis, Fig.~\ref{XRD_FWHM} compares the full width half maximum (FWHM) of the (110), (020) and (001) XRD peaks as a function of the \textit{a}-axis length. It is clear that the change in the FWHM of the (110) peak amounts 300\%, much stronger than those of  the other peaks (40\% for (020) and 16\% for (001)). This contrast indicates that the broadening is intimately related to the degree of dimer formation since the dimer-bond direction is predominantly along the \textit{a} axis, but not to the degree of mosaicity which would broaden all peaks equally.

The cross section of the X-ray scattering is sensitive to the atomic number (\textit{Z}). Since both lithium (\textit{Z }= 3) and oxygen (\textit{Z}=8) are light elements, conventional XRD used in this study does not contain much of their information. To confirm this, we performed simulations of  XRD patterns (not shown) changing the structure factor of the O ($Z=8$) using the space group $P2_1/m$. The relative intensity of the peaks (110), (020) and (001) are found to remain invariant despite structure factor is changed. This fact confirms that these peaks in \lro\ are sensitive only to Ru atoms. Thus the broadening of those peaks is related to the Ru positions.

The difference in sharpness of the peaks is attributed to the spatial coherence of the short bond (dimer), which is mainly aligned along the \textit{a}-axis with a smaller component along the \textit{b}-axis (see Fig.~\ref{cell_init}(d)); non-dimerized Ru ions result in decoherence of the Ru dimers, leading to broader peaks. Thus, the present results indicate that the degree of formation of the Ru dimers is crucially dependent on the sample preparation process. 

This scenario is supported by the variation in the obtained lattice parameters, which are summarized in Table~\ref{cellpar}. It is clear that the variations of the cell parameters have a definite tendency. The strongest sample dependence among the four lattice parameters is seen in the \textit{a} parameter (nearly 0.41\%). Shorter \textit{a} for well treated samples indicates well-ordered Ru-Ru dimers. In contrast, the changes in \textit{b}, \textit{c} and $\beta$ are only 0.19, 0.06\% and 0.07\%, respectively. It is a bit surprising that samples with well-ordered dimers (such as G and H) have longer \textit{b}, in contrast to shorter \textit{a} in such samples. Possible scenarios for this systematic enlargement of \textit{b} when \textit{a} is shorter are proposed in Sec.~\ref{magnetization}.

\subsection{\label{magnetization} Magnetic behavior and its correlation with the crystalline structure}

\begin{figure}[tb]
\includegraphics[width=9cm]{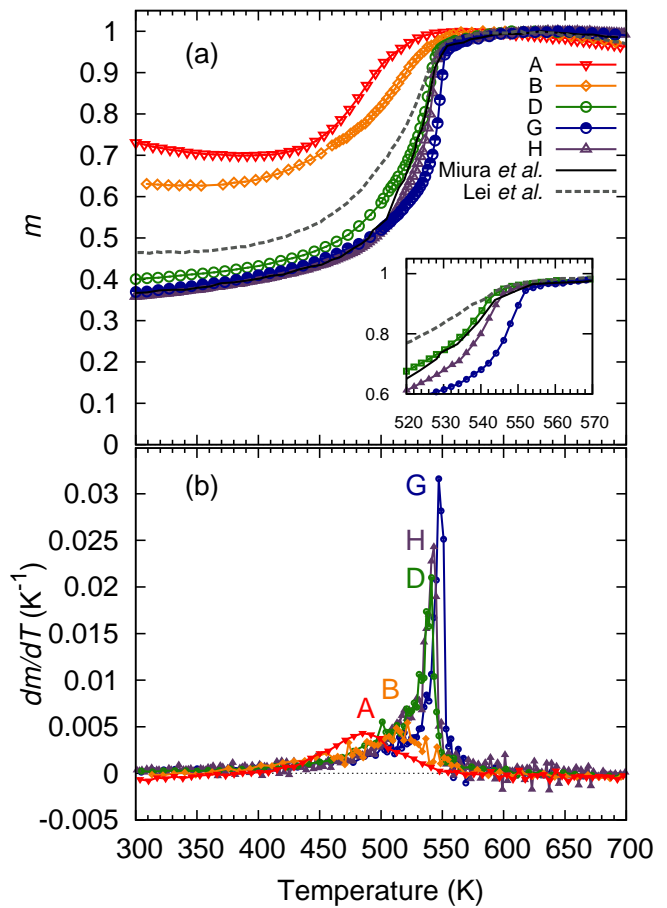}
\caption{\label{MT_dmdt} (Color online) (a) Temperature dependence of the magnetization ratio $m\equiv M/M_{\text{max}}$ at 10 kOe of representative samples. Results taken from Miura \textit{et al.} \cite{miura_new-type_2007} and Lei \textit{et al.} \cite{lei_structural_2014} are included for comparison. Further analysis cannot be performed since the method and devices for the measurement may differ from ours. Inset shows the behavior near the transition. The diamagnetism due to the core electrons has been subtracted \cite{bain_diamagnetic_2008}. (b) Temperature derivative of $m$.}
\end{figure}

In Fig.~\ref{MT_dmdt}(a), we present the temperature dependence of the magnetization at 10 kOe normalized by the maximum magnetization $m\equiv M$/$M_{\text{max}}$ ($M_{\text{max}}~\sim M_{610~\text{K}}$ depending on samples) of representative samples. In this plot, diamagnetic contributions of ion cores \cite{bain_diamagnetic_2008} have been subtracted. Similar to the lattice parameters, the magnetic behavior is strongly sample dependent. In particular, the magnetic transition is very broad for samples A and B, prepared with only one heating process. In contrast, for samples G and H, which have been prepared by more elaborate processes, a jump-like behavior in $m(T)$ at $T_\text{d}$ is observed. As shown in the inset of Fig.~\ref{MT_dmdt}(a), sample G exhibits a sharp onset at 553~K, noticeably higher than the previously reported $T_\text{d}$ of 540~K \cite{miura_new-type_2007}. The difference in the sharpness of the magnetic transitions is more evident in the temperature derivative of $m(T)$ plotted in Fig.~\ref{MT_dmdt}(b). For samples G and H, $dm/dT$ has a much sharper peak with large height and smaller width compared to other samples.
The jump in $m(T)$ and the sharp behavior in $dm/dT$ for samples G and H indicate that the dimer transition is intrinsically the first-order transition, with a finite jump in the magnetization for an ideal case. This is compatible with a recent study where the dimer transition is found to be the first-order transition by means of differential thermal analysis (DTA) \cite{terasaki_ruthenium_2015}.

One comment should be made on the high-temperature crystal structure above $T_{\text{d}}$. Initially, the space group at high temperature was proposed to be  $C2/m$ \cite{miura_new-type_2007}. However, it was mentioned in that paper that the Rietveld refinement of neutron diffraction is equally good for the space groups $C2/m$ and $C2/c$. The space group $C2/m$  was chosen under the premise that the dimer transition is a second-order one. Since the transition is now found to be first order, the possibilities of the space group $C2/c$ as well as $C2/m$ for the high temperature phase need to be reconsidered on equal footing \cite{miura_new-type_2007}.  We should note that this assignment is for the averaged ``thermodynamic'' structure which consists of dynamical changes of dimer configurations.

Sharp changes of physical quantities at a first-order transition can be rounded by the effect of inhomogeneities (disorder) and the discontinuity can be completely vanished when the amount of disorder reaches a certain limit, as explained in Ref.~\cite{imry_influence_1979}. Based on the rounded curves of samples A and B (Fig.~\ref{MT_dmdt}(a)) and the heating treatment of those samples (Table~\ref{cellpar}), it is suggestive that the rounding of the transitions is caused by disorder. We confirmed that measuring several times from 300 K to 700 K and vice versa does not affect the sharpness of the transition or the value of the  magnetization at 300~K (not shown). This indicates that the disorder is retained even in the high temperature phase (at least up to 700~K). The possible distortions will be discussed in the final part of this section.

\begin{figure}[tb]
\includegraphics[width=9cm]{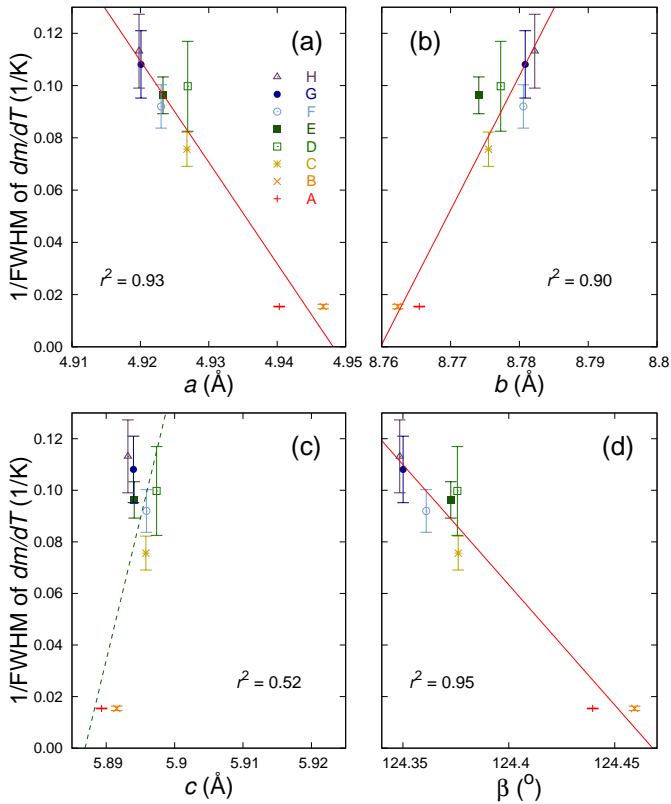}
\caption{\label{FWHM} (Color online) Comparison between the sharpness of the magnetic transition and various structural parameters. The former is characterized by the inverse of the full with at half maximum (FWHM) of \textit{dm/dT}(\textit{T}). We plot this quantity as functions of the cell parameters \textit{a, b, c} and $\beta$ in (a--d). In (a--c) the same horizontal scales (0.04 \AA) for the full scale are used to highlight the differences in the change of the lattice parameters. The red and green lines are linear fittings whose square of the Pearson correlation coefficients ($r^2$) are shown in these panels. Alphabets A, B, C, etc. correspond to the sample labels in Table~\ref{cellpar}. }
\end{figure}

Given these data, it is important to investigate relations between the crystal structure and magnetic behavior. We plot in Fig.~\ref{FWHM} the inverse of the FWHM of the peak in $dm/dT$, as functions of several lattice parameters. The former characterizes the sharpness of the magnetic transition. 
According to the values of the square of the Pearson correlation coefficients ($r^2$) shown in Fig.~\ref{FWHM}, the main correlation between the structural parameters and sharpness of the magnetic transition is present in the cell parameters \textit{a}, \textit{b}  and $\beta$, but not in \textit{c} (see the definitions of the cell parameters in Fig.~\ref{cell_init}).

\begin{figure}[tb]
\includegraphics[width=9cm]{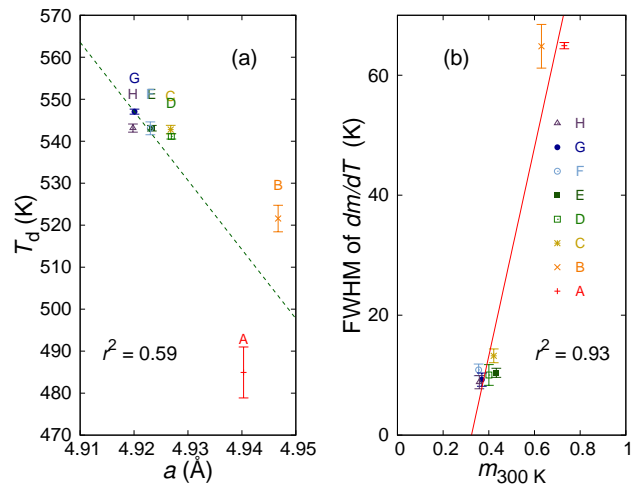}
\caption{\label{Td_M300K} (Color online) (a) Relation between $T_\text{d}$ determined by the peak of $dm/dT (T)$ and the lattice parameter $a$. (b) Correlation between the sharpness of the magnetic transition and $m_{300~\text{K}}$. The diamagnetic contributions of the ionic cores have been substracted \cite{bain_diamagnetic_2008}. The red line indicates linear correlation while the green dashed line indicate weak linear correlation.}
\end{figure}

We have chosen the parameter \textit{a} for quantifying the in-plane distortion for hexagonal lattices and examine relations to behavior at the magnetic transition. As seen in Fig.~\ref{FWHM}(a), samples whose \textit{a} value is larger have a broader magnetic transition, whereas samples whose \textit{a} is shorter have a sharper transition (i.e. samples F, G and H). The temperature of the magnetic transition, $T_{\text{d}}$, is also related to the lattice parameters. Here, $T_{\text{d}}$ is determined by the maximum point in $dm/dT$. In Fig.~\ref{Td_M300K}(a), we present $T_{\text{d}}$ plotted against \textit{a}. Samples exhibiting shorter \textit{a} tend to have higher $T_{\text{d}}$.

In addition to the sharpness of the magnetic transitions, values of the normalized magnetization at 300~K $m_{300~\text{K}}=M_{300~\text{K}}/M_{\text{max}} $ also exhibit correlation with some other properties. In Fig.~\ref{Td_M300K}(b), we show the relation between the FWHM of $dm/dT$ and $m_{300~\text{K}}$.  From the linear fitting, we can extrapolate to the ideal case of FWHM~$\rightarrow0$ to yield $m_{300~\text{K}} = 0.33~\pm~0.02$. The measured value of the maximum susceptibility of sample H after the diamagnetic corrections is $\chi_{\text{max}}= 5.04 \times 10^{-4}$~emu/mol at 610~K. 
By using this ideal ratio $m_{300~\text{K}}$ and $\chi_{\text{max}}$ for sample H, we deduce for an ideal dimer structure the magnetization at room temperature to be $\chi_{300\text{ K}}\sim 1.68\times10^{-4}$ emu/mol.
So far this large residual magnetization at room temperature has been attributed to the Van Vleck paramagnetism \cite{miura_new-type_2007, jackeli_classical_2008, khaliullin_excitonic_2013}, but not confirmed yet.

\begin{figure}[tb]
\includegraphics[width=8cm]{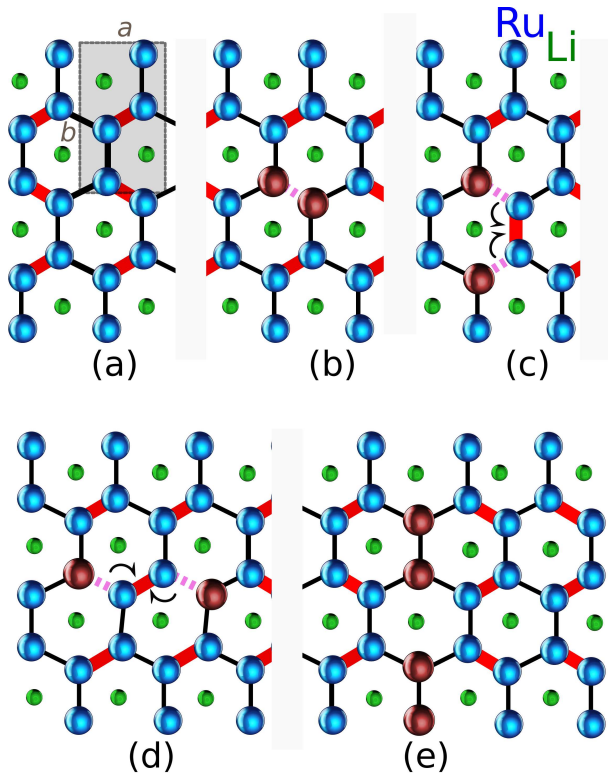}
\caption{\label{cell} (Color online) Schematic of the Ru honeycomb layer of Li$_2$RuO$_3$ with Ru-Ru dimers. Red and black lines indicate the short (dimer) and long bonds. Dashed pink lines indicate the positions where dimer bonds are located in a perfect dimerized structure. Blue and red spheres represent dimerized and non-dimerized Ru ions. A unit cell \cite{miura_new-type_2007} in the \textit{ab} plane, as well as the cell parameters \textit{a} and \textit{b}, is also shown. (a) Ideal dimer pattern with ``\textit{a}-axis dimer''. (b) With a pair of non-dimerized Ru-ions. (c) With a dimer in the \textit{b} direction. (d) With a misplaced ``\textit{a}-axis dimer''. (e) With a boundary of dimer domains.}
\end{figure}

Here, we discuss the origin of the observed coupling between 
the lattice and magnetic features.
As already described, the sample dependence of lattice properties
can be explained by the coherence of the dimer formation.
The variation in the magnetic transition is also related
to the coherence of the dimer ordering.
Since the spins are expected to form non-magnetic spin-singlet within a dimer \cite{miura_new-type_2007,kimber_valence_2014}, 
the reduction of magnetization represents the amount of Ru ions participating in the dimer formation.
Overall, these results indicate that the coherent formation of the Ru dimers is the origin of the magneto-lattice coupling.

We further discuss possible magnetic and crystalline defects responsible for the broadening of the (\textit{h}00) peaks, elongation of the parameter \textit{a}, shortening of the parameter \textit{b}, and the broadening of the magnetic transition. Naively, an unpaired Ru ion, represented by a red sphere in Fig.~\ref{cell}(b), is accompanied by an active magnetic moment. Such non-dimerized ions should contribute to elongate the cell parameter \textit{a} as well as the broadening of the peaks with $h\neq0$  compared with the ideal dimer phase.
Since unpaired ions have higher energy as pointed out in Ref.~\cite{kimber_valence_2014}, formation of dimers along the \textit{b} direction (Fig.~\ref{cell}(c)) should also be considered.
Indeed, the energy required to form the \textit{b}-axis dimer is expected to be similar to the energy of the other dimers \cite{kimber_valence_2014}, which we shall call the ``\textit{a}-axis dimer''.
Such dimers should be accompanied by unpaired ions with active spins, increasing the magnetization at room temperature ($m_{300 \text{K}}$).
Formation of the \textit{b}-axis dimer may also explain shrinkage of the \textit{b}-axis length for samples A and B. In addition, misplaced ``\textit{a}-axis dimers'' (Fig.~\ref{cell}(d)) also result in the broadening of the $h\neq0$ peaks of XRD, accompanying unpaired Ru ions that contribute to the enhancement of magnetization.
Another possible defect is domain walls between regions with opposite dimer patterns, as shown in Fig.~\ref{cell}(e).
Note that, in each hexagon, the four Ru bonds having the \textit{a}-axis component are equivalent in the high-temperature phase~\cite{miura_new-type_2007}.
Thus, formation of two different domains is plausible.
Along the domain wall, either \textit{b}-axis dimers or unpaired Ru ions exist. So far it is an open and interesting question which defects is most dominant in actual samples. 

We note that the broadening of XRD peaks attributable to stacking faults has been previously reported in Li$_2$MnO$_3$, which has a similar honeycomb structure \cite{breger_high-resolution_2005}. The broadening of the XRD peaks in \lro\ (Fig.~\ref{XRD_pristine}) might be affected by the stacking faults as well. However, one important difference between Li$_2$MnO$_3$ and \lro\ is the presence of dimers with short Ru-Ru bonds in the ruthenate,  in which the amount of in-plane distortion $\sqrt{3}a/b$
and the sliding angle $\beta$ are much greater. In this paper, we have demonstrated that the broadening of the XRD peaks and the broadening of the magnetic transition associated with dimerization are intimately related. Because stacking faults alone cannot explain the observed magnetic behavior, the dimer decoherence in each layer must be involved in the XRD peak broadening.

As we already mentioned, decoherence and defects in the dimer formation are triggered by another kind of defects already imprinted in the high-temperature structure.
According to the previous thermogravimetric analysis (TGA) of \lro\ \cite{omalley_structural_2009}, this compound is stable up to $\sim 1200$\celsius\ without any detectable loss in the mass. On the other hand, we observe increment of the amount of impurity RuO$_2$ phase while improving the dimer coherence by repeating the grinding and heating process. This segregation of RuO$_2$ implies that Li is deficient in samples with less coherent dimers, probably due to the evaporation of Li$_2$CO$_3$ in the initial heating. Then, samples with coherent dimers are formed after heating again to 1000\celsius\ by rearranging Li and releasing RuO$_2$.
In \lro, there are two Li sites in the crystal structure: the inter-plane site and the in-plane site at the center of the Ru hexagon. Since Li at the inter-plane site is expected to be more movable and defects in this site should be relaxed at lower temperatures, we infer that defects in the in-plane site are more likely to be responsible for the decoherence of the Ru dimer formation. To verify this scenario, studies of the intentional introduction of Li defects may be valuable. We should note that the dimer coherence can be indirectly disturbed also by stacking faults, since they affects the interlayer coupling of the dimers. 

Before closing this section, we compare our results with those in previous studies. The broad magnetic transition of samples with less coherent dimers such as samples A and B (Fig.~\ref{MT_dmdt}) resembles those reported for samples partially substituted with Ir or Ti \cite{lei_structural_2014, terasaki_ruthenium_2015}. This fact raises a possibility that the effects on the dimer transition are mainly caused through the randomness introduced by isovalent substitution, rather than changes in the electronic structure. Similarly, the randomness effect on the absence of the dimer transition in single crystals reported in Ref.~\cite{wang_lattice-tuned_2014} should be re-examined.

\section{\label{CONCLUSION}CONCLUSION}
We have examined crystalline and magnetic properties of various Li$_2$RuO$_3$ samples. In order to obtain high-quality polycrystalline samples exhibiting intrinsic coherent dimer formation in Li$_2$RuO$_3$, the following elements are important: more than one grinding/heating process, final heating at 1000\celsius , and thorough homogenization before heating to reduce RuO$_2$ impurities.
In fact, by improving the sample dimer coherence, we show that the intrinsic magnetic transition is of first order and has a sharp onset at as high as 553~K, noticeably higher than 540~K reported previously. 
It has been demonstrated that disorder sensitively affects the coherence of dimerization, which is reflected systematically in the lattice parameters as well as the magnetic transition. We expect these results provide important bases for elucidating the intrinsic properties of Li$_2$RuO$_3$. 
 
\section{\label{ACKNOWLEDGMENTS}ACKNOWLEDGMENTS}
We acknowledge useful discussions with G. Khaliullin. This work was supported by Grant-in-Aid for Scientific Research (KAKENHI~26247060) from Japan Society for the Promotion of Science (JSPS) and Grant-in-Aid for Scientific Research on Innovative Areas "Topological Materials Science" (KAKENHI 15H05852) from the Ministry of Education, Culture, Sports, Science and Technology (MEXT). Marco-Polo Jimenez-Segura is also supported by MEXT.


\input{reference.bbl}

\end{document}

%% file: reference.bbl
%